\documentstyle[11pt,newpasp,twoside]{article}
\def\edcomment#1{\iffalse\marginpar{\raggedright\sl#1\/}\else\relax\fi}
\reversemarginpar

\begin{document}
\title{Far-Ultraviolet Spectra of AGN: First Results from FUSE}
\author{Gerard A. Kriss}
\affil{Space Telescope Science Institute, 3700 San Martin Drive,
Baltimore, MD 21218}

\begin{abstract}
Composite spectra derived from HST observations of moderate redshift ($z\sim 1$)
quasars show that the peak of the spectral energy distribution is at
$\sim$1000 \AA, and that the ionizing continuum falls off smoothly to
shorter wavelengths as $f_\nu \sim \nu^{-1.8}$.
HUT and FUSE observations of nearby AGN (e.g., 3C 273) show similar results.
FUSE spectra reveal a wealth of detail in the
912--1200 \AA\ spectral range, including multi-component intrinsic absorption
in the {\sc O~vi} doublet and Ly$\beta$, and surprisingly strong narrow-line
emission from {\sc O~vi} $\lambda\lambda 1032,1038$.
Monitoring observations of AGN below 1200 \AA\ are feasible, and such
experiments could help to unravel the structure of intrinsic absorbers in AGN
as well as the highest ionization portions of the broad-line region.
\end{abstract}

\section{Introduction}

The far-ultraviolet wavelength range spans the peak of the intrinsic
spectral energy distribution of active galactic nuclei (AGN).
Thus, it is important for understanding the energy generation mechanism
and the processes that govern accretion onto massive black holes.
At low to moderate redshift, the low-energy portion of the ionizing continuum
enters the far-UV spectral range, and it can be observed with spectrographs
on board the {\it Hubble Space Telescope} (HST) and instruments such as
the {\it Hopkins Ultraviolet Telescope} (HUT) (Davidsen et al. 1992)
and the {\it Far Ultraviolet Spectroscopic Explorer} (FUSE) (Moos et al. 2000).
Since this portion of the spectrum determines the radiative input to the
broad-line region (BLR) and the narrow-line region (NLR) in AGN,
as well as the surrounding host galaxy and the intergalactic medium (IGM),
determining the spectral shape in individual objects as well as on average
is a crucial input for understanding the physical conditions of the gas
in surrounding regions.

The 900--1200 \AA\ spectral range contains numerous diagnostic spectral
features that can be applied to AGN physics.
The most prominent emission line is the {\sc O~vi} resonance doublet
$\lambda\lambda 1032,1038$, which, as shown in Fig. 1, is particularly
strong in the spectra of low-redshift, low-luminosity AGN.
This line can serve as a diagnostic of the energy input from the extreme
ultraviolet to soft X-ray portions of the ionizing continuum.
Likewise, the {\sc O~vi} doublet is a crucial diagnostic for the
warm absorbing gas commonly seen as {\sc O~vii} and {\sc O~viii} absorption
in X-ray spectra of AGN (Reynolds 1997; George et al. 1998).
The high-order Lyman lines and the Lyman limit provide additional
diagnostics of absorbing gas.
In some cases (e.g., NGC 4151 or NGC 3516) the neutral hydrogen can be
optically thick and thereby play a significant role in collimating the
ionizing radiation that illuminates the NLR (Kriss et al. 1997).
Finally, numerous ground-state transitions of molecular hydrogen
in the Lyman and Werner bands provide a sensitive tracer of molecular gas.
Under the right circumstances, one might expect to see $\rm H_2$ associated
with the obscuring torus in AGN in absorption against the continuum and
broad emission lines.

\begin{figure}
\plotfiddle{"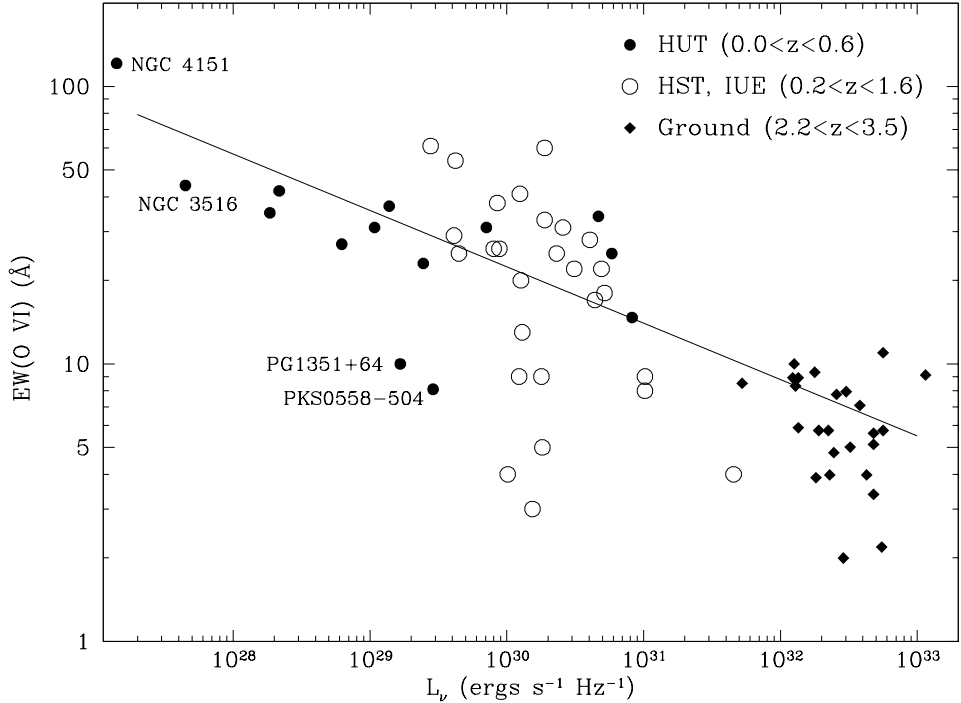"}{3.5in}{-90}{50}{50}{-197}{285}
\caption{
Low-luminosity, low-redshift AGN have the relatively strongest O VI emission
seen in AGN. (Updated from Zheng, Kriss, \& Davidsen 1995).
}
\end{figure}

\section{The Ionizing Continuum Shape}

The ionizing continuum shape in AGN is difficult to study directly.
At low redshift it is mostly obscured by our own interstellar medium, and
inferences must be made from the spectral shape observed in the far-UV and
soft X-ray (on either side of the opaque interstellar absorption) as well
as from clues provided by the emission lines observed.
The notion that continuum radiation of AGN peaked in the unobservable
extreme ultraviolet can be traced back to the accretion-disk fits of
Malkan \& Sargent (1982) as well as the ``big blue bump" suggested by the
UV and soft X-ray excesses of PG1211+143 (Bechtold et al. 1987).
Accounting for the fluxes in the broad emission lines led Mathews \& Ferland
(1987) to adopt an ionizing spectral shape peaking in the extreme ultraviolet
(see Fig. 2).

\begin{figure}
\plotfiddle{"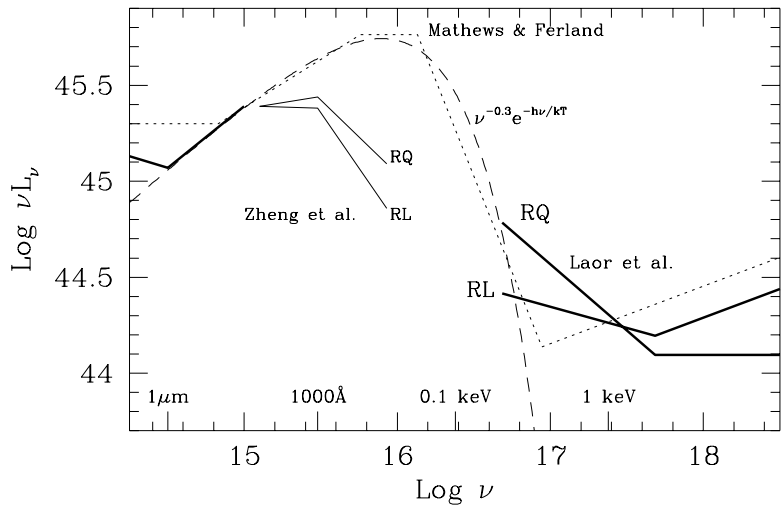"}{2.5in}{0}{70}{70}{-200}{-256}
\caption{
Schematic representations of inferred spectral energy distributions of AGN
after Laor et al. (1997).  The $\nu^{-0.3}e^{-h\nu/kT}$ curve approximates
a simple accretion disk spectrum.
}
\end{figure}

More recent work that uses ensembles of objects to produce a ``composite"
spectrum over the observable wavelength range suggests that there is no
extreme ultraviolet peak.
The UV composites of Zheng et al. (1997) for radio-loud (RL) and radio-quiet
(RQ) objects, shown schematically in Fig. 2, have a blue far-UV continuum with
a distinct break to a steeper spectral index ($f_\nu \sim \nu^{-1.8}$)
at $\sim$1000 \AA.
Soft X-ray spectral composites constructed by Laor et al. (1997) have
slopes similar to the short wavelength end of the UV composites.

The Zheng et al. (1997) composite builds upon the earlier examples of
Baldwin (1977) and Francis et al. (1992) to produce an average QSO spectrum
from HST ultraviolet spectra of 100 different QSOs.
This composite consists of all FOS observations of QSOs with $z > 0.33$
prior to 1996 January 1.
To construct the composite, the observed fluxes were first corrected for
foreground Galactic reddening with a Seaton (1978) law using extinction as
specified by Burstein \& Heiles (1978).
Strong Lyman $\alpha$ lines and visible Lyman-limit systems were corrected
in the individual spectra. (Objects with strong intrinsic absorption such as
BALQSOs were simply omitted from the composite.)
Fainter Ly$\alpha$ forest lines and the ``Lyman valley"
(M{\o}ller \& Jakobsen 1990) were corrected for statistically
as a function of redshift.
The result from Zheng et al. (1997) is shown in Fig. 3.

\begin{figure}
\plotfiddle{"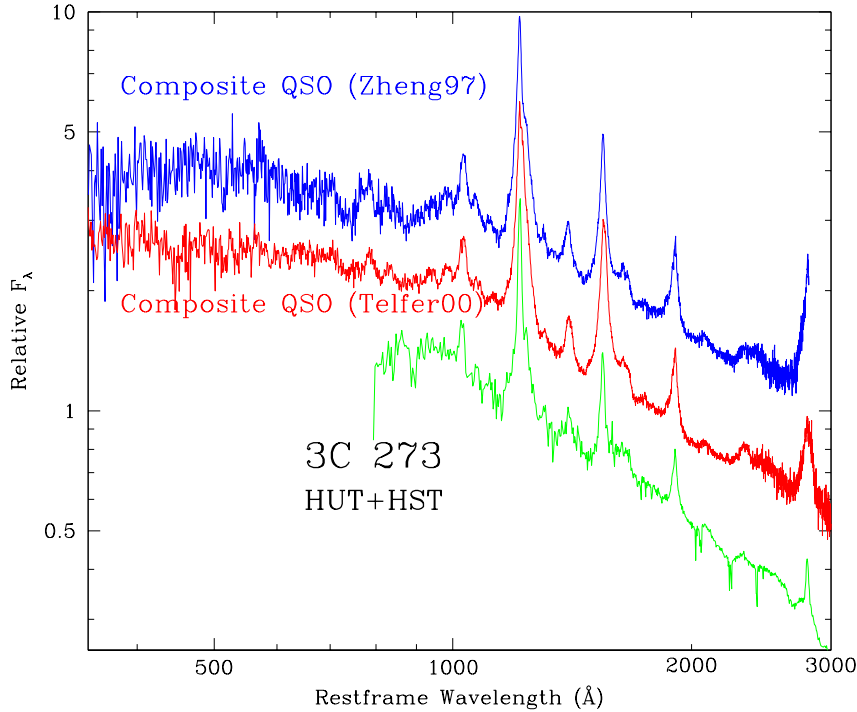"}{3.53in}{-90}{50}{50}{-197}{285}
\caption{
The composite QSO spectrum of Zheng et al. (1997) shows a break in spectral
index at 1000 \AA\ where it flattens to $f_\nu \sim \nu^{-1.7}$.
An updated composite using more than two times as many quasars (Telfer et al. 2000) shows the same characteristics.  The far-UV spectrum of 3C 273 is
qualitatively similar to both---the spectrum breaks at 1030 \AA\ to an
index of $1.7 \pm 0.36$.
}
\end{figure}

Work currently in progress by Telfer et al. (2000) is updating this
composite spectrum using the same technique with a sample that is nearly
three times larger.  Containing spectra of 248 QSOs obtained prior to
2000 May 31, the Telfer et al. (2000) composite is nearly identical
to the Zheng et al. (1997) result.
A physical interpretation of the spectral shape observed by Zheng et al. (1997)
is that the typical AGN spectrum is Comptonized radiation from an accretion
disk.
A spectral break near the Lyman limit is a characteristic expected in
Comptonized accretion-disk spectra.
Comptonization smears out the Lyman-limit edge of the intrinsic accretion-disk
spectrum and produces a power-law tail
at wavelengths shortward of $\sim 1000$~\AA.
Although a simple
model of a Comptonized disk provides a good fit to the Zheng et al. composite,
there is the nagging question of how physical such a composite is,
and whether its shape holds any physical meaning since the individual spectra
of the sample QSOs vary in shape over quite a wide range.
The study of the 3C~273 far-UV spectrum by Kriss et al. (1999) provides a firm
physical basis for the interpretation of the Zheng et al. composite.
In the 3C~273 spectrum (also shown in Fig. 3), one sees the same
qualitative characteristics as in the composite
spectrum---the spectral index steepens at $\sim$1000 \AA\ in the rest frame.
As shown further in Fig. 4, the shorter-wavelength UV
spectrum extrapolates very well to the observed soft-X-ray spectrum.
Both the composite spectra and the spectrum of 3C~273 leave little
room in the extreme ultraviolet for a big bump in the QSO spectral energy
distribution.

\begin{figure}
\plotfiddle{"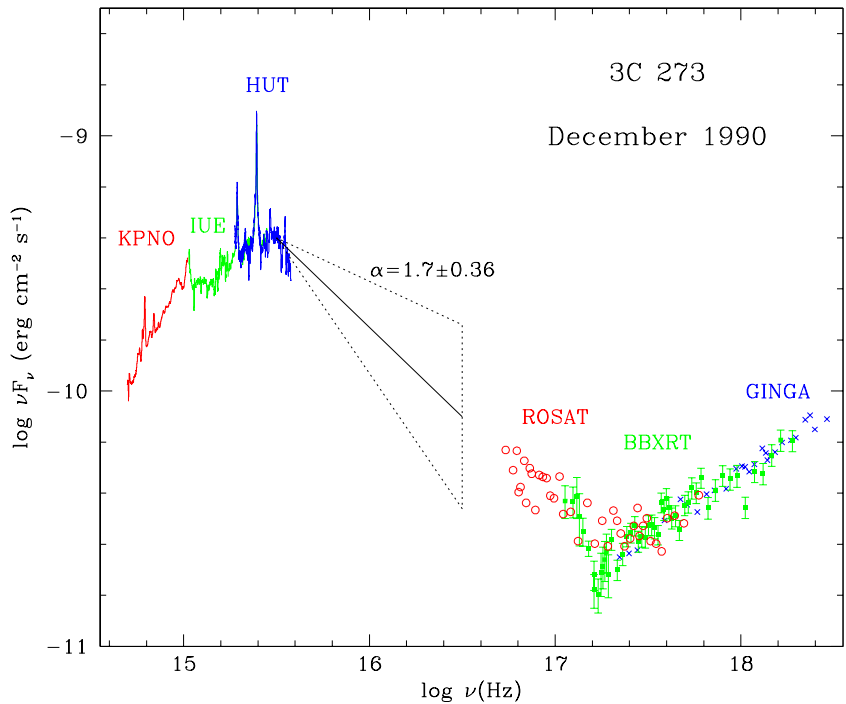"}{3.5in}{-90}{50}{50}{-197}{285}
\caption{
The far-UV and soft X-ray spectra of 3C 273 allow little room for a bump in the extreme ultraviolet.
}
\end{figure}

The apparent lack of an extreme ultraviolet peak in the ionizing spectrum now
leaves us in a quandary.  The analysis of Mathews \& Ferland (1987) based on
the broad emission line fluxes is still valid; Korista, Ferland, \&
Baldwin (1997) suggest that the BLR ``sees" a different continuum shape
than we do since the He~{\sc ii} $\lambda$1640 equivalent widths in
moderate redshift QSOs require more ionizing photons than are present in the
Zheng et al. composite.
On the other hand, the mean spectral index of the QSO composite matches the
ionization requirements for {\sc H~i} and He~{\sc ii} absorption by the IGM
(Zhang et al. 1997), so apparently the IGM is seeing the same average that
we are.
It may be that the answer lies in detailed studies of individual objects.
Objects with high He~{\sc ii} equivalent widths may dominate the He~{\sc ii}
average, but these same objects may not be represented in the sample of spectra
contributing to the composite at short wavelengths.
Far-UV and soft X-ray spectra of objects at moderate redshift that tightly
constrain the possible EUV spectral shape (as shown above for 3C~273) may
resolve this issue.

\section{The FUSE AGN Program}

\begin{figure}
\plotfiddle{"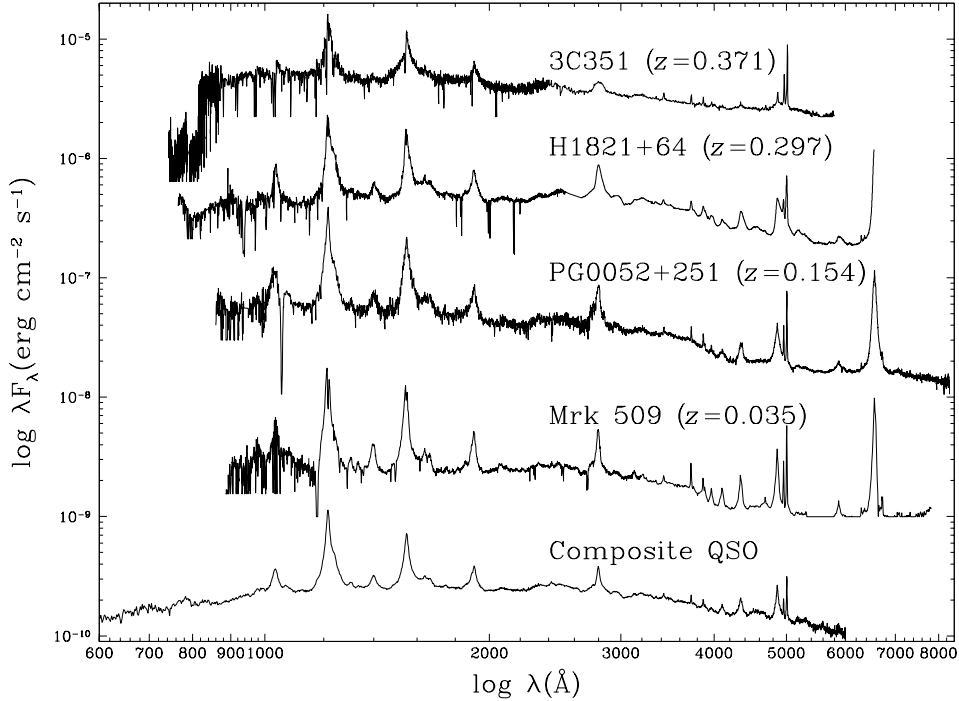"}{3.5in}{-90}{50}{50}{-197}{285}
\caption{
Combined FUSE, HST, and ground-based data for four AGN show spectral energy
distributions peaking at $\sim$1000 \AA, in qualitative similarity to the
composite QSO spectrum of Zheng et al. (1997).
The sharp drop at the short wavelength end of the 3C 351 spectrum is a
foreground Lyman-limit system at $z=0.22$.
}
\end{figure}

The launch of the Far Ultraviolet Spectroscopic Explorer (FUSE) on 1999 June 24
provided a new tool in orbit for studying the far-UV spectra of AGN.
FUSE observations cover the 905--1187 \AA\ band with a resolving power
of $\sim 20,000$.  The mission concept is described by Moos et al. (2000),
and the initial in-flight performance is presented by Sahnow et al. (2000).
Basically, four separate primary mirrors gather light for
four prime-focus, Rowland-circle spectrographs.
Two of the optical systems employ LiF coatings to cover the 990--1187 \AA\ band,
and the other two use SiC coatings to get reflectivity down to
wavelengths as short as 905 \AA.
Holographically ruled gratings disperse the light entering through the
selected entrance slit (a 30\arcsec$\times$30\arcsec\ aperture for most
observations) and form an astigmatic image on the two-dimensional
photon-counting microchannel-plate detectors.
The detectors have KBr photocathodes and delay-line anode readouts
that provide the location and arrival time of each photon event.

\begin{figure}
\plotfiddle{"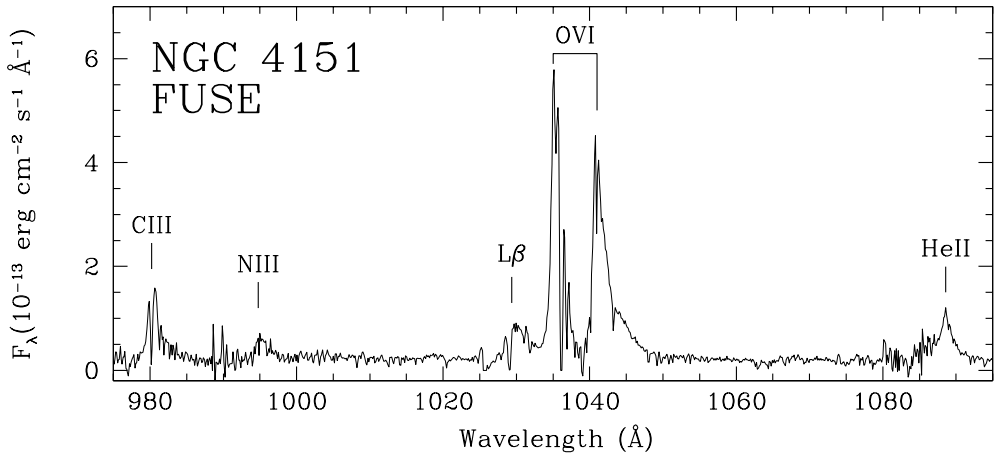"}{2.13in}{-90}{62}{62}{-240}{345}
\caption{
The FUSE spectrum of NGC 4151 shows a continuum flux $20 \times$ lower than
that observed with HUT in 1995 (Kriss et al. 1995). The FUSE spectrum is
dominated by high excitation narrow-line emission, and the optically thick
blanket of broad Lyman-line absorption is no longer apparent.
}
\end{figure}

FUSE began routine observations in December 1999, and here I will present
the first results from the PI-team program directed toward the study of AGN.
This program piggy-backs off data acquired as part of the larger PI-team
study of {\sc O~vi} absorption in the Galactic halo and the study of the
deuterium to hydrogen ratio along different sight lines through the
Galactic halo and high-velocity clouds.  As part of these programs,
the $\sim100$ UV-brightest AGN are being surveyed at low to moderate
signal-to-noise (S/N) ratios to determine the best sight lines for further
study.  A few intrinsically interesting AGN are being observed at high S/N
as part of the ``medium-size" team project as are a few Seyfert 2 galaxies.
In addition to the FUSE observations, an HST snapshot program is obtaining
contemporaneous spectra at longer UV wavelengths using the STIS G140L and
G230L gratings, and current-epoch ground-based optical spectra covering
$\sim$3500--9000 \AA\ are being obtained by R. Green and M. Brotherton at KPNO.
The primary scientific goals of the FUSE AGN program are to
(1) determine the far-UV continuum shape of low-redshift AGN;
(2) examine the strength and profile of the broad {\sc O~vi} emission line;
(3) search for intrinsic {\sc O~vi} and Lyman-limit absorbers and study
their kinematics;
(4) search for intrinsic $\rm H_2$ absorption that may be associated with the
obscuring torus; and
(5) study the strengths of other far-UV resonance lines such as {\sc C~iii}
$\lambda 977$ and {\sc N~iii} $\lambda 991$.

As of 2000 July 1 we have observed 47 AGN at $z < 1$.
46 of these are Type 1 AGN;
one is the Seyfert 2 Mrk~463, which was not detected.
Of the 28 Type 1s with $z < 0.15$
(so that {\sc O~vi} falls in the FUSE bandpass),
strong, broad {\sc O~vi} emission is visible in all but one (Pks $0558-504$).
Four of the 28 also show strong {\it narrow} {\sc O~vi} emission lines.
Approximately 40\% (11/28) show intrinsic {\sc O~vi} absorption.
No intrinsic Lyman limits are detected, and no intrinsic molecular
hydrogen absorption is visible.
Representative spectral energy distributions for 4 of the AGN in this sample are
shown in Fig. 5.
Note the overall qualitative similarity of these AGN spectra to the
Zheng et al. (1997) composite.
While it is too early to draw definitive conclusions, this suggests that
there is little evolution in mean spectral shape from redshifts of $\sim$1,
as represented by the composite, to the current epoch.

\section{FUSE Observations of NGC 4151}

\begin{figure}
\plotfiddle{"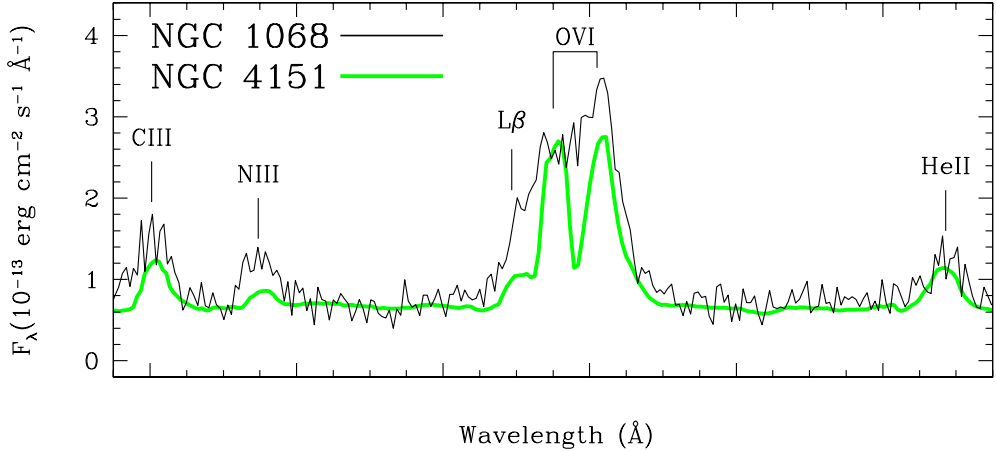"}{2.13in}{-90}{62}{62}{-240}{198}
\caption{
If the FUSE spectrum of NGC 4151 is smoothed to $\sim3$~\AA\ resolution,
it is nearly identical to the HUT spectrum of NGC 1068 (Kriss et al. 1992b).
}
\end{figure}

Turning to a more detailed discussion of individual objects observed by FUSE,
let me start with the spectrum of a recurrent favorite: NGC~4151.
This FUSE observation was obtained as part of a coordinated set
that included {\it Chandra} grating spectra and HST/STIS echelle spectra.
While our primary intent was to study the rich absorption-line spectrum of
NGC~4151, we found to our surprise that NGC~4151 was $\sim20 \times$ fainter
than in recent HUT (Kriss et al. 1995) and ORFEUS (Espey et al. 1998)
observations.
As shown in Fig. 6, the FUSE spectrum is dominated by narrow-line emission,
rather than the usual bright continuum and broad {\sc O~vi} seen in this
spectral range.  The usual broad absorption lines, including the opaque
Lyman limit, are not readily apparent.
The narrow lines have a full-width-half-maximum of $\sim450~\rm km~s^{-1}$,
and they are at the systemic velocity of the galaxy.
Some residual broad {\sc O~vi} emission can be seen at the base of the
narrow lines.

The HUT and ORFEUS observations of NGC~4151 showed that the Lyman absorption
lines were broad and saturated, yet their relative equivalent widths
suggested that they only partially covered the broad-line and continuum
emission, with $10-20$\% of the flux leaking through or around (due to
scattering) the absorbing material.
The broad absorption that is present in the {\sc O~vi} region is not
very deep in the FUSE spectrum, and its profile never drops below the current
continuum flux level.
This suggests that the little continuum emission we do see
($\sim 3 \times 10^{-14}~\rm erg~cm^{-2}~s^{-1}~\AA^{-1}$ at 1030 \AA)
may be completely dominated by scattered light, and that the current
low flux state of NGC~4151 is not due to intrinsic dimming of the
central engine, but rather heavier-than-usual obscuration that is completely
opaque at UV wavelengths.
Note that the red and blue-shifted ``satellite" lines often visible around
{\sc C~iv} $\lambda1549$ in earlier low states of NGC~4151 (Ulrich et al.
1985; Clavel et al. 1987) do not appear to have any counterparts in the
current {\sc O~vi} profile.  (See the paper by M. Crenshaw in this
volume for a possible explanation of the {\sc C~iv} satellite lines.)

\begin{figure}
\plotfiddle{"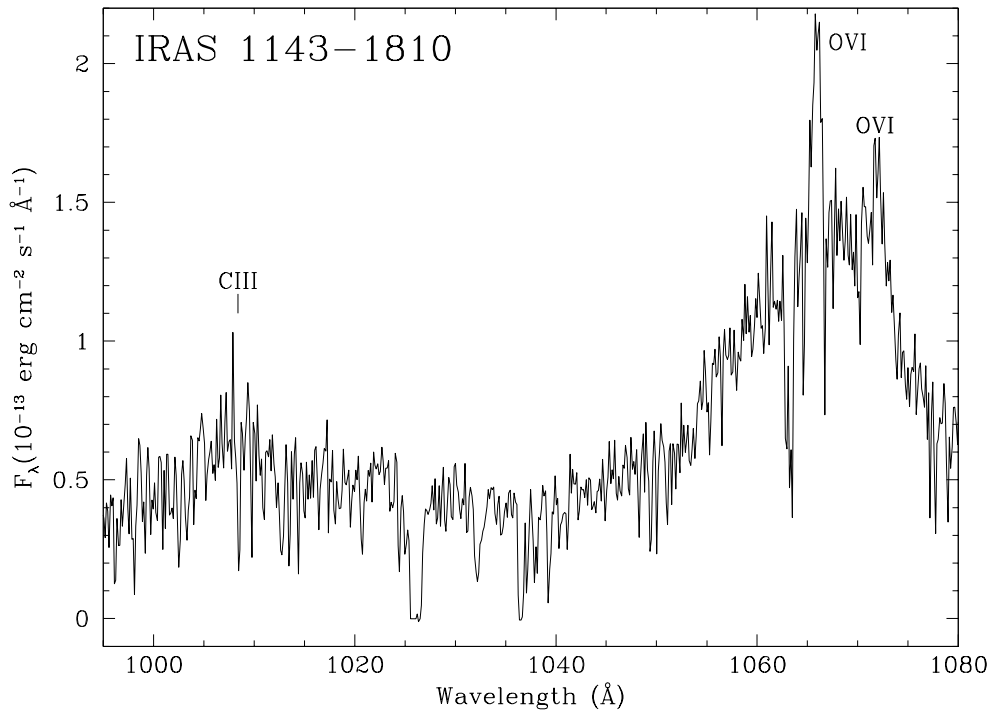"}{3.5in}{-90}{50}{50}{-197}{285}
\caption{
The FUSE spectrum of IRAS $1143-1483$ shows strong, {\it narrow} O~VI emission
superposed on a broad O VI emission line.  Broad  C~III $\lambda 977$ is also
prominent.
}
\end{figure}

The prominent narrow lines we see in the spectrum are very reminiscent of
the HUT spectrum of NGC~1068, where the high excitation lines of
{\sc C~iii} $\lambda977$ and {\sc N~iii} $\lambda991$ were particularly
prominent (Kriss et al. 1992b; Grimes, Kriss, \& Espey 1999).
In Fig. 7 we have used a Gaussian kernel to smooth the FUSE spectrum
of NGC~4151 down to the $\sim3$ \AA\ resolution of HUT and overlayed
it directly on the HUT spectrum of NGC~1068.
The correspondence is quite remarkable.
Whether one thinks the narrow-line emission is powered predominantly by
kinetic energy in shocks or photoionization by the central engine, it
is clear that the same mechanism is at work in the NLRs of both NGC~1068
and NGC~4151.
This provides even more support for unified models of Seyfert 1s and Seyfert 2s
based on the obscuration, orientation, and reflection paradigm
(Antonucci 1993).

\section{Narrow {\sc O~vi} Lines in FUSE AGN Spectra}

NGC~4151 is not the only AGN to show prominent narrow emission lines in its
FUSE spectrum.  So far, strong narrow {\sc O~vi} emission lines have been
detected in 4 out of 28 objects with $z < 0.15$: NGC~4151, NGC~3516, NGC~5548,
and IRAS $1143-1483$.
Such a high percentage is a bit surprising since strong narrow-line
emission in lower-excitation UV lines such as {\sc C~iv} $\lambda1549$
is rarely seen in HST spectra.
The one well-known exception is the HST/FOS spectrum of NGC~5548 in a low
state (Crenshaw, Boggess, \& Wu 1993).
The FUSE observations suggest that there is a very high ionization component
of the NLR present in many AGN.

The physical conditions in this region are unclear.
While in three of the present cases they are easily seen because the galaxy
is in a low flux state (NGC~4151, NGC~3516, and NGC~5548), as shown in Fig. 8
for IRAS $1143-1483$, the narrow {\sc O~vi} components are quite visible
superposed on a strong broad {\sc O~vi} profile.
In three of these cases the narrow lines also have the usual 2:1 flux ratios
indicative of optically thin gas.
However, as shown in Fig. 9, the lines in NGC~3516 have a 1:1 ratio more
indicative of optically thick gas.

\begin{figure}
\plotfiddle{"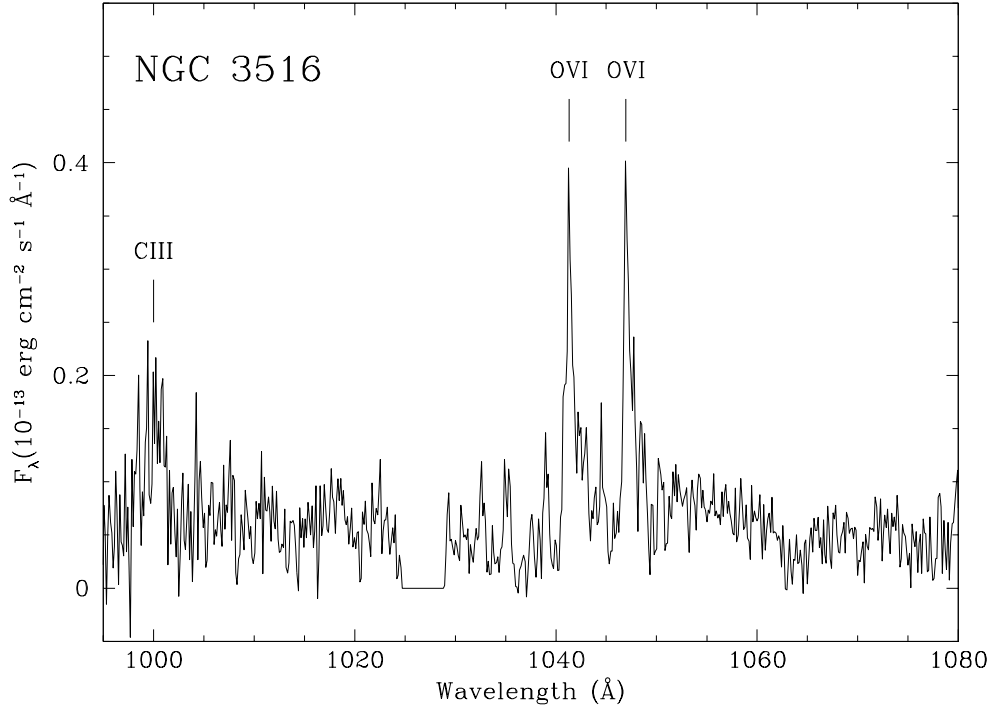"}{3.5in}{-90}{50}{50}{-197}{285}
\caption{
The FUSE observation of NGC~3516 shows strong, narrow O~VI emission lines.
As evidenced by the 1:1 intensity ratio, these lines are optically thick.
}
\end{figure}

\section{FUSE Observations of Warm Absorbers}

The ``warm" (or, ionized) absorbing gas that is common in the X-ray spectra
of low-redshift AGN is a major new component of their near-nuclear structure.
About half of all low-redshift AGN show absorption
by ionized gas (Reynolds 1997; George et al. 1998), and a similar fraction show
associated UV absorption in highly ionized species such as C IV (Crenshaw et al.
1999).
For AGN that have been observed at moderate to high spectral resolution in
both the X-ray and the UV, there is a one-to-one correspondence between
objects showing X-ray and UV absorption, suggesting that the phenomena are
related in some way (Crenshaw et al. 1999).
The gas has a total mass exceeding $\sim 10^3~\rm M_\odot$ (greater than the
broad-line region, or BLR), and is outflowing at a rate
$>0.1~\rm M_\odot~yr^{-1}$
($10 \times$ the accretion rate in some objects) (Reynolds et al. 1997).

The role played by the ionized absorbing gas in AGN at the moment is unclear.
The UV-absorbing gas lies exterior to the BLR, as revealed by the depth of
the absorption lines in many cases (Crenshaw et al. 1999).
This gas could either arise in winds from the molecular torus
(Weymann et al. 1991; Emmering et al. 1992), the accretion disk
(K\"onigl \& Kartje 1994; Murray et al. 1995), or
stripped stellar envelopes in the BLR (Alexander \& Netzer 1994;
Scoville \& Norman 1996).
Or, the X-ray heated winds that provide the reflecting medium in Type 2 AGN
(Krolik \& Begelman 1986) may be seen as the warm absorbing medium in Type 1
AGN (Krolik \& Kriss 1995).
Understanding the absorbing gas in nearby AGN may
help us to understand how radiation is collimated (Kriss et al. 1997).
This goes not only to the heart of understanding the structure
of AGN, but is also a crucial element in understanding the
ionizing radiation field at high redshifts and links between
AGN and the X-ray background.
If the obscuration/orientation paradigm for AGN unification is
correct, then the solid angle subtended by unobscured lines of
sight from an active nucleus is a crucial parameter in determining
the fraction of ionizing radiation that illuminates the surrounding IGM.
Obscured Type 2 AGN may also comprise a significant fraction
of the point sources producing the X-ray background.
Understanding the intrinsic structures that lead to the
Type 1/Type 2 distinction is a vital clue in our overall
understanding of these phenomena.

A key question for understanding warm absorbers is
how the X-ray and UV spectral domains are related.
In some cases, UV absorbing gas may be directly associated with the X-ray
warm absorber
(3C351: Mathur et al. 1994; NGC 5548: Mathur et al. 1995).
In other cases, however, the UV gas appears to be in an even
lower ionization state, and there is no direct relation between the
X-ray absorption and the UV absorption (NGC 4151: Kriss et al. 1995;
NGC 3516: Kriss et al. 1996a, 1996b;
NGC 7469: Kriss et al. 2000a).
High spectral resolution observations of {\sc O~vi}
absorption in AGN with X-ray warm absorbers is proving to be an invaluable
tool for definitively analyzing these issues.

The disparity between the UV and X-ray absorbers is illustrated most clearly
by FUSE observations of the Seyfert 1 galaxy Mrk 509 (Kriss et al. 2000b).
As shown in Fig. 10, this observation resolves the UV absorption
into seven kinematic components. In addition to the {\sc O~vi} and Ly$\beta$
absorption shown here, {\sc C~iii} $\lambda977$ lines are seen in the
lowest-ionization components, and Lyman-line absorption up to Ly$\zeta$
permits an accurate measure of the neutral hydrogen column density and its
covering fraction. Comparison of the {\sc O~vi}, {\sc H~i},
and {\sc C~iii} column densities to photoionization models
permits one to independently assess the total
column density of each kinematic component.
Table 1 summarizes the physical properties of each of the seven components.
Of the seven systems present, only \#5, a system near the
systemic velocity of Mrk 509, stands out clearly as having a level of
ionization and total column density compatible with an X-ray warm absorber.
Note, however, that the total equivalent width of the UV absorption is
dominated by the other six components, yet the X-ray absorption that
would be associated with them is negligible.

The {\sc O~vi} and {\sc O~vi} edges seen by Reynolds (1997) in ASCA spectra
of Mrk~509 imply column densities for the ionized gas of
$N_{O7} = (4.6^{+1.3}_{-1.7}) \times 10^{17}~\rm cm^{-2}$ and
$N_{O8} = (3.7^{+3.7}_{-2.8}) \times 10^{17}~\rm cm^{-2}$.
The photoionization model for the UV-absorbing component \#5 computed by
Kriss et al. predicts
$N_{O7} = 2.3 \times 10^{17}~\rm cm^{-2}$ and
$N_{O8} = 0.1 \times 10^{17}~\rm cm^{-2}$.
Given the large uncertainties in the X-ray columns and the temporal
difference between the X-ray and UV observations,
the agreement is remarkably close, and it is quite likely that
component \#5 is same gas as that absorbing the X-ray radiation.
It is puzzling, however, that this component lies so close to the systemic
velocity.  Most absorbers in other AGN (including most of the other components
in this one) are blue-shifted relative to the host galaxy, and this is true
for the high-resolution X-ray spectra obtained so far with {\it Chandra}:
the X-ray absorption lines in NGC~5548 are blue-shifted by
$\sim280~\rm km~s^{-1}$ (Kaastra et al. 2000) and by $\sim440~\rm km~s^{-1}$
in NGC~3783 (Kaspi et al. 2000).
Clearly a larger sample of objects that would permit us to assess these
differences is essential.

\begin{figure}
\plotfiddle{"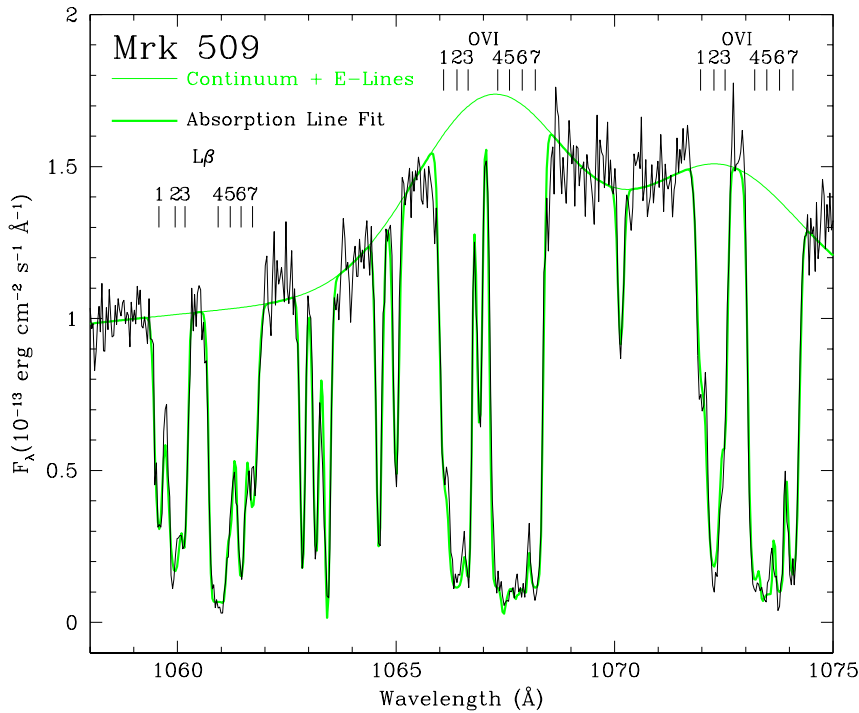"}{3.6in}{-90}{50}{50}{-197}{285}
\caption{
The FUSE observation of Mrk 509 shows intrinsic absorption in O VI and Ly$\beta$
with a total of 7 kinematic components (Kriss et al. 2000b).
}
\end{figure}

\begin{table}
\begin{center}
\caption{Physical Properties of the Absorbers in Mrk~509\label{tbl-2}}
\begin{tabular}{l c c c c c c}
\tableline
{\#} &
{$\Delta \rm v$} &
{$\rm N_{OVI}$} &
{$\rm N_{HI}$} &
{$\rm N_{OVI} / N_{HI}$} &
{$\rm N_{tot}$} &
{log U} \\
{ } &
{$\rm ( km~s^{-1} )$} &
{$\rm ( cm^{-2} )$} &
{$\rm ( cm^{-2} )$} &
{ } &
{$\rm ( cm^{-2} )$} &
{ } \\
\tableline
1 & $-438$ & $1.9\times10^{14}$ & $5.7\times10^{14}$ & \phantom{0}0.37 & $2.0\times10^{18}$ & $-1.64$ \\
2 & $-349$ & $1.3\times10^{15}$ & $5.3\times10^{14}$ & \phantom{0}1.51 & $5.9\times10^{18}$ & $-1.19$ \\
3 & $-280$ & $1.2\times10^{14}$ & $9.3\times10^{14}$ & \phantom{0}0.48 & $2.2\times10^{18}$ & $-1.79$ \\
4 & \phantom{0}$-75$ & $1.2\times10^{15}$ & $6.0\times10^{15}$ & \phantom{0}0.19 & $1.6\times10^{19}$ & $-1.73$ \\
5 & \phantom{00}$-5$  & $3.2\times10^{15}$ & $2.7\times10^{14}$ & 13.9  & $5.0\times10^{20}$ & $-0.43$ \\
6 & \phantom{0}$+71$ & $1.2\times10^{15}$ & $1.2\times10^{15}$ & \phantom{0}2.14 & $7.6\times10^{18}$ & $-1.41$ \\
7 & $+166$ & $9.4\times10^{14}$ & $1.2\times10^{15}$ & \phantom{0}2.76 & $6.8\times10^{18}$ & $-1.46$ \\
\tableline
\tableline
\end{tabular}
\end{center}
\end{table}

\begin{figure}
\plotfiddle{"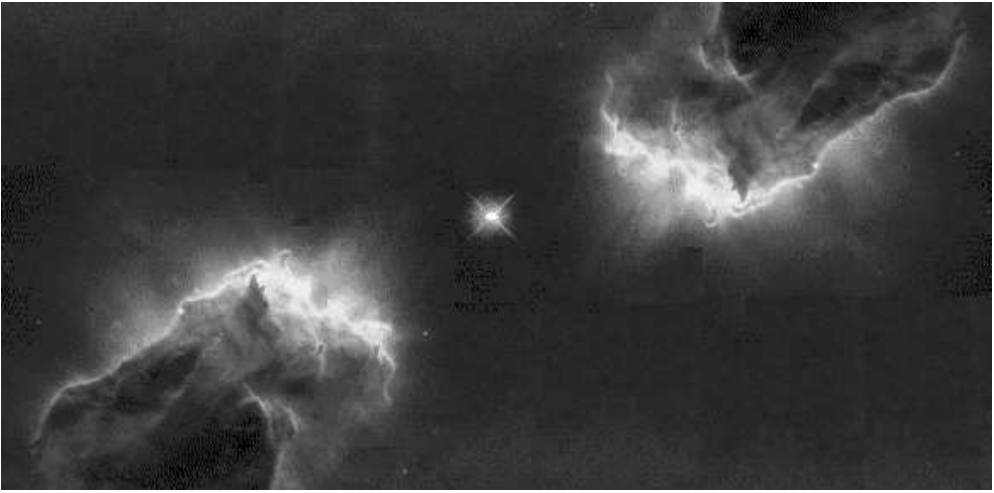"}{2.6in}{0}{64}{64}{-196}{-150}
\caption{
Artistic rendering of how a molecular torus surrounding an AGN might appear
based on HST images of the Eagle Nebula.
}
\end{figure}

The multiple kinematic components frequently seen in the UV absorption spectra
of AGN clearly show that the absorbing medium is complex, with separate
UV and X-ray dominant zones.  One potential geometry is high density,
low column UV-absorbing clouds embedded in a low density,
high ionization medium that dominates the X-ray absorption.
This is possibly a wind driven off the obscuring torus or the accretion disk.
What would this look like in reality?
Detailed modeling of dense molecular gas irradiated by the ionizing continuum
of an AGN is likely to miss the complexities that arise as material is
ablated from the surface and flows away.
We will not soon get a close-up look at this aspect of an AGN, so it is
instructive to look at nearby analogies.
The HST images of the pillars of gas in the Eagle Nebula, M16,
show the wealth of detailed structure in gas evaporated from a molecular
cloud by the UV radiation of nearby newly formed stars (Hester et al. 1996).
Fig. 11 shows what this might look like in an AGN---a dense molecular torus
surrounded by blobs, wisps, and filaments of gas at various densities.
It is plausible that the multiple UV absorption lines seen in AGN with
warm absorbers are caused by high-density blobs of gas embedded in a hotter,
more tenuous, surrounding medium, which is itself responsible for the X-ray
absorption.  Higher density blobs would have lower ionization
parameters, and their small size would account for the low overall column
densities.

At sight lines close to the surface of the obscuring torus, one might
expect to see some absorption due to molecular hydrogen.
Given the dominance of Type 1 AGN in our observations so far, the lack of
any intrinsic $\rm H_2$ absorption is not too surprising since our
sight lines are probably far above the obscuring torus.
NGC~4151 and NGC~3516 are examples where the inclination may be more favorable
since these objects have shown optically thick Lyman limits in the past
(Kriss et al. 1992; Kriss et al. 1995; Kriss et al. 1996a), but our
FUSE observations do not show such high levels of neutral hydrogen
at the current epoch.
Molecular hydrogen will not survive long in an environment with a strong
UV flux.  Given this, one might think that any sightline in an AGN in which
a strong UV continuum was visible could not show $\rm H_2$ absorption
since the $\rm H_2$ would not be optically thick enough to be self shielding.
However, if the $\rm H_2$ is contained in small, dense blobs that have
ablated from the obscuring torus, these blobs could be thick enough to
shield their cores and small enough that they do not completely obscure
the central radiation source.
Thus, the $\rm H_2$ absorption, if present, is likely to show up as
optically thick, high-column-density components that only partially cover the
continuum or broad lines.
Our best prospects for seeing such material are in upcoming FUSE observations
of NGC~1068.

As I've shown here, FUSE offers numerous advantages for the study of
absorbing gas in AGN.
First, since the Ly$\alpha$ absorption in the HST bandpass is often saturated,
it is difficult to deblend the various
velocity components and obtain accurate {\sc H~i} column densities.
The unsaturated high-order Lyman lines in the FUSE bandpass make accurate
measurement of the {\sc H~i} column density straightforward.
Second, the {\sc O~vi} resonance lines are a key link between the
lower-ionization species seen in the UV and the higher-ionization gas observed
with {\it Chandra}.
The presence of different, adjacent ionization states in the two bandpasses
makes it possible to ascertain the ionization structure in a more
model-independent way. One need not make assumptions about abundances or use
photoionization models, for example, as would be required in comparing
{\sc C~iv} in the HST band to {\sc O~vii} and {\sc O~viii} in the {\it Chandra}
spectrum.
Similarly, the ionization state of the UV-absorbing gas can be obtained
more directly without recourse to assumptions about abundances by comparing
the {\sc C~iii} opacity in the FUSE bandpass at 977 \AA\ to {\sc C~iv} as
observed with HST.
Finally, after using {\sc O~vi} to identify the kinematic components in the
absorbing gas associated with the X-ray warm absorber, the FUSE observations
can be used to trace the kinematics of the X-ray absorbing gas at higher
velocity resolution than is possible with the {\it Chandra}
grating observations.

\section{Monitoring Low-Redshift AGN Below 1200 \AA}

Not much is known about the variability of nearby AGN in the 912--1200 \AA\ 
bandpass since observations at these wavelengths have been sparse.
The {\it Astro} and {\it ORFEUS} shuttle missions have provided some
insights, but FUSE promises to give the first real opportunity to study
short and longer term variability in this band.

To date, the only repetitive observations of an AGN below 1200 \AA\ have
been the HUT and ORFEUS observations of NGC~4151 (Kriss et al. 1992a;
Kriss et al. 1995; Kriss et al. 1997; Espey et al. 1998).
These showed that broad {\sc O~vi} emission responds immediately to continuum
changes, to within a resolution of 1--2 days.
The saturated broad absorption lines such as Ly$\beta$ and {\sc O~vi} show
little response to continuum changes, presumably because the changes in column
density required to produce noticeable changes in equivalent width are larger
than those induced by the varying continuum.
However, weaker absorption features such as {\sc C~iii} $\lambda 1176$ and the
high-order Lyman series (which is optically thick at the Lyman limit in
these observations of NGC~4151) respond to continuum changes with a delayed
response of $\sim$5 days, reflecting the recombination timescale of the
photoionized gas. Observations proposed with FUSE to monitor the Seyfert~1
galaxy NGC~3783 contemporaneously with HST and {\it Chandra} grating
observations could give the first glimpse of the response of the gas in
a warm absorber to changes in the ionizing continuum at both X-ray and
far-UV wavelengths.

Potential future monitoring experiments such as {\it Kronos} could profit from
an ability to observe in the 912--1200 \AA\ spectral range.
As noted earlier, {\sc O~vi} is bright in low-redshift AGN; it is nearly as
bright as {\sc C~iv}.
While it probes a zone in the BLR similar in ionization to He~{\sc ii},
it is much brighter than He~{\sc ii} $\lambda 1640$.
For studies of warm absorbers, the {\sc O~vi} lines and the higher-order
Lyman lines offer several advantages, as noted earlier.

Observing in this bandpass requires some observational trade-offs, however.
Good throughput requires LiF coatings on the optics and open-window detectors.
Both of these features entail significant costs in development and in
contamination management throughout the instrument development and
integration process.
Sealed detectors and $\rm MgF_2$ coatings are easier to manage, and their
use can improve the throughput at longer UV wavelengths (e.g., {\sc C~iv}
$\lambda 1549$) by up to a factor of 2.

\section{Summary}

I've reviewed the continuum, emission-line, and absorption-line properties
of primarily low-redshift AGN, concentrating on the first observations
available from FUSE.
Composite far-UV spectra of QSOs developed from a sample now consisting of
over 200 AGN (Telfer et al. 2000) confirms the result of Zheng et al.
(1997) that the continuum spectral index breaks at $\sim1000$ \AA\ to a
steep power law of spectral index $\sim 1.8$.
Observations of individual objects such as 3C~273 (Kriss et al. 1999)
show a qualitatively similar feature. 
Initial results from FUSE (with contemporaneous HST and ground-based
observations) also show that the spectral energy distributions of low-redshift
AGN are peaking in the 1000--1200 \AA\ range.
This implies that there is little evolution in the mean shape of the
ionizing continuum from $z > 1$ to the current epoch.

Nearly all Type 1 AGN observed with FUSE show strong, broad {\sc O~vi}
emission.  The broad {\sc O~vi} emission of AGN as revealed by HUT and
FUSE is significantly stronger in low-redshift objects.
The FUSE observations also show that strong, {\it narrow} {\sc O~vi} emission
is prominent in several Type 1 AGN.
In most cases this appears to be emission from the narrow-line region
made relatively more prominent because the continuum and broad lines are
in a relatively low state.
However, NGC~3516 presents an interesting (and not yet understood) exception
since the narrow {\sc O~vi} doublet has an optically thick 1:1 intensity ratio.

FUSE observations of {\sc O~vi} and high-order Lyman-line absorption in
AGN with warm absorbers offers a powerful new tool for the study of the
warm absorbing gas.  The FUSE resolution reveals multiple kinematic
components, and the {\sc O~vi} absorption lines often identify the
highest ionization system likely to be directly associated with the
warm absorber. Proposed monitoring observations may help to pinpoint the
location and physical conditions in this important structural component
of the nuclear region in AGN.

\acknowledgments
The FUSE results are due in large part to the help of the FUSE Team and the
FUSE AGN Working Group, particularly M. Brotherton and W. Zheng.
R. Telfer, J. Kim, and A. Koratkar also made significant contributions.
This work was supported in part by NASA contract NAS5-32985 for the FUSE
project, NASA Long Term Space Astrophysics grant NAGW-4443, and grant
GO-08144.01-97A from the Space Telescope Science Institute,
which is operated by the Association of Universities for Research in Astronomy,
Inc., under NASA contract NAS5-26555.


\begin{references}
\reference Alexander, T., \& Netzer, H. 1994, \mnras, 270, 781
\reference Antonucci, R. 1993, \araa, 31, 473
\reference Baldwin, J. 1977, \mnras, 178, 67P
\reference Bechtold, J., Czerny, B., Elvis, M., Fabbiano, G., \&
	Green, R. F. 1987, \apj, 314, 699
\reference Burstein, D., \& Heiles, C. 1978, \apj, 225, 40
\reference Clavel, J., et al. 1987, \apj, 321, 251
\reference Crenshaw. D. M., Boggess, A., \& Wu, C.-C. 1993, \apj, 416, L67
\reference Crenshaw. D. M., Kraemer, S. B., Boggess, A., Maran, S. P.,
	Mushotzky, R. F., \& Wu, C. C. 1999, \apj, 516, 750
\reference Davidsen et al. 1992, \apj, 392, 264
\reference Emmering, R. T., Blandford, R. D., \& Shlossman, I.  1992, 
	\apj, 385, 460
\reference Espey, B.R., Kriss, G.A., Krolik, J. H., Zheng, W.,
	Tsvetanov, Z., \& Davidsen, A.F. 1998, \apj, 500, L13
\reference Francis, P., Hewett, P. C., Foltz, C., Chaffee, F. H.,
	Weymann, R. J., \& Morris, S. L. 1991, \apj, 373, 465
\reference George, I. M., Turner, T. J., Netzer, H., Nandra, K.,
	Mushotzky, R. F., \& Yaqoob, T. 1998, \apjs, 114, 73
\reference Grimes, J. P., Kriss, G. A., \& Espey, B. R. 1999, \apj, 526, 130
\reference Hester, J. J., et al. 1996, AJ, 111, 2349
\reference Kaastra, J. S., Mewe, R., Liedahl, D. A., Komossa, S.,
	\& Brinkman, A. C. 2000, \aap, 354, L931
\reference Kaspi, S., Brandt, W. N., Netzer, H., Sambruna, R.,
	Chartas, G., Garmire, G. P., \& Nousek, J. A. 2000, \apj, 535, L17
\reference K\"onigl, A., \& Kartje, J. F. 1994, \apj, 434, 446
\reference Korista, K., Ferland, G., \& Baldwin, J. 1987, \apj, 487, 555
\reference Krolik, J. H., \& Begelman, M. C. 1986, \apj, 308, L55
\reference Krolik, J. H., \& Kriss, G. A. 1995, \apj, 447, 512
\reference Kriss, G. A., Davidsen, A. F., Blair, W. P., Ferguson, H. C., \&
	Long, K. S. 1992b, \apj, 394, L37
\reference Kriss, G. A., Davidsen, A. F., Zheng, W., Kruk, J. W., \&
	Espey, B. R.  1995, \apj, 454, L7
\reference Kriss, G. A., Davidsen, A. F., Zheng, W., \&
	Lee, G. 1999, \apj, 528, 123
\reference Kriss, G. A., et al. 1992a, \apj, 392, 485
\reference Kriss G.~A., et al. 1996a, \apj, 467, 622
\reference Kriss G.~A., et al. 1996b, \apj, 467, 629
\reference Kriss, G. A., Peterson, B. M., Crenshaw, D. M., \& Zheng, W. 2000a,
	\apj, 535, 58
\reference Kriss, G. A., et al. 2000b, \apj, 538, L17
\reference Kriss, G., Krolik, J., Grimes, J., Tsvetanov, Z.,
	Zheng, W., \& Davidsen, A.  1997, in ASP Conf. Ser. Vol. 113,
	Emission Lines in Active Galaxies: New Methods and Techniques, eds.
	B.M. Peterson, F.-Z. Cheng, \& A.S. Wilson
	(San Francisco: ASP), 453
\reference Krolik, J. H., \& Kriss, G. A. 1995, \apj, 447, 512
\reference Laor, A., et al. 1997, \apj, 477, 93
\reference Malkan, M. A., \& Sargent, W. L. W., 1982, \apj, 254, 22
\reference Mathews, W. G., \& Ferland, G. J. 1987, \apj, 323, 456
\reference Mathur, S., Wilkes, B., \& Elvis, M. 1995, \apj, 452, 230
\reference Mathur, S., Wilkes, B., Elvis, M., \& Fiore, F. 1994, \apj, 434, 493
\reference M{\o}ller, P., \& Jakobsen, P. 1990, \aap, 228, 299
\reference Moos, W., et al. 2000, \apj, 538, L1
\reference Murray, N., Chiang, J., Grossman, S. A., Voigt, G. M. 1995,
	\apj, 451, 498
\reference Nandra, K., \& Pounds, K. 1994, \mnras, 268, 405
\reference Reynolds, C. S. 1997, \mnras, 286, 513
\reference Sahnow, D., et al. 2000, \apj, 538, L7
\reference Seaton, M. 1979, \mnras, 187, 73P
\reference Scoville, N., \& Norman, C. 1995, \apj, 451, 510
\reference Telfer, R. C., Kriss, G. A., Zheng, W., \& Davidsen, A. F. 2000,
	in preparation
\reference Ulrich, M.-H. 1985, Nature, 313, 745
\reference Weymann, R. J., Morris, S. L., Foltz, C. B., \& Hewett, P. C. 1991,
	\apj, 373, 23
\reference Zhang et al. 1997, \apj, 485, 496
\reference Zheng, W., Kriss, G. A., \& Davidsen, A. F. 1995, \apj, 440, 606
\reference Zheng, W., Kriss, G. A., Telfer, R. C., Grimes, J. P., \& Davidsen,
	A. F. 1997, \apj, 475, 469
\end{references}
\end{document}